\documentclass[aps,twocolumn,prb]{revtex4-1}
\usepackage{amsmath}
\usepackage{natbib}
\usepackage[pdftex]{graphicx}

\usepackage[usenames]{color}

\newcommand{\R}[1] {{\color{red} #1}}

\begin{document}
\date{\today}
\author{Brankica Jankovic, Jeannette Ruf, Claudio Zanobini, Olga Bozovic, David Buhrke, Peter Hamm$^*$}
\affiliation{Department of Chemistry, University of Zurich, Winterthurerstr.\ 190, CH-8057 Z\"urich, Switzerland\\$^*$peter.hamm@chem.uzh.ch}

\title {Sequence of Events During Peptide Unbinding from RNase S: A Complete Experimental Description}

\begin{abstract}  \textbf{Abstract}: The photo-triggered unbinding of the intrinsically disordered S-peptide from the RNase~S complex is studied with the help of transient IR spectroscopy, covering a wide range of time scales from 100~ps to 10~ms.
To that end, an azobenzene moiety has been linked to the S-peptide in a way that its helicity is disrupted by light, thereby initiating its complete unbinding. The full sequence of events is observed, starting from unfolding of the helical structure of the  S-peptide on a 20~ns timescale while still being in the binding pocket of the S-protein, S-peptide unbinding after 300~$\mu$s, and the structural response of the S-protein after 3~ms. With regard to the S-peptide dynamics, the binding mechanism can be classified as an induced fit, while the structural response of the S-protein is better described as conformational selection.\\

\centering{\R{TOC Graphics:}}

\centering\includegraphics[width=0.4\textwidth]{FigTOC.pdf}

\end{abstract}
\maketitle

Investigation of protein-protein interactions takes one of the central places in protein biochemistry and biophysics. Proteins are team players that can exert their numerous biological roles only through the intertwined interaction networks with other partners. Clarification of the structural and dynamical aspects underlying protein/ligand binding mechanisms represents a problem of fundamental importance.\cite{boehr2009, morando2016} Understanding this phenomenon helps explaining the structure-function paradigm and greatly facilitates the \textit {de novo} design of proteins with new, predictable functions.\cite{boehr2009} The current understanding of protein binding mechanisms is based on the two limiting scenarios – induced fit and conformational selection.\cite{boehr2009, csermely2010, morando2016} The classification rests on the sequence of binding events vs conformational changes.\cite{vogt2012, gianni2014, hammes2009} An emerging view considers the induced fit as an extreme case of conformational selection, and is defined more widely as conformational selection with additional conformational adjustment.\cite{csermely2010}
This becomes particularly important for intrinsically disordered proteins, which adopt very diverse conformations during their interaction with different binding partners.\cite{dyson2005, wright2015, wright2009}

\begin{figure}[t]
	\centering
	\includegraphics[width=0.4\textwidth]{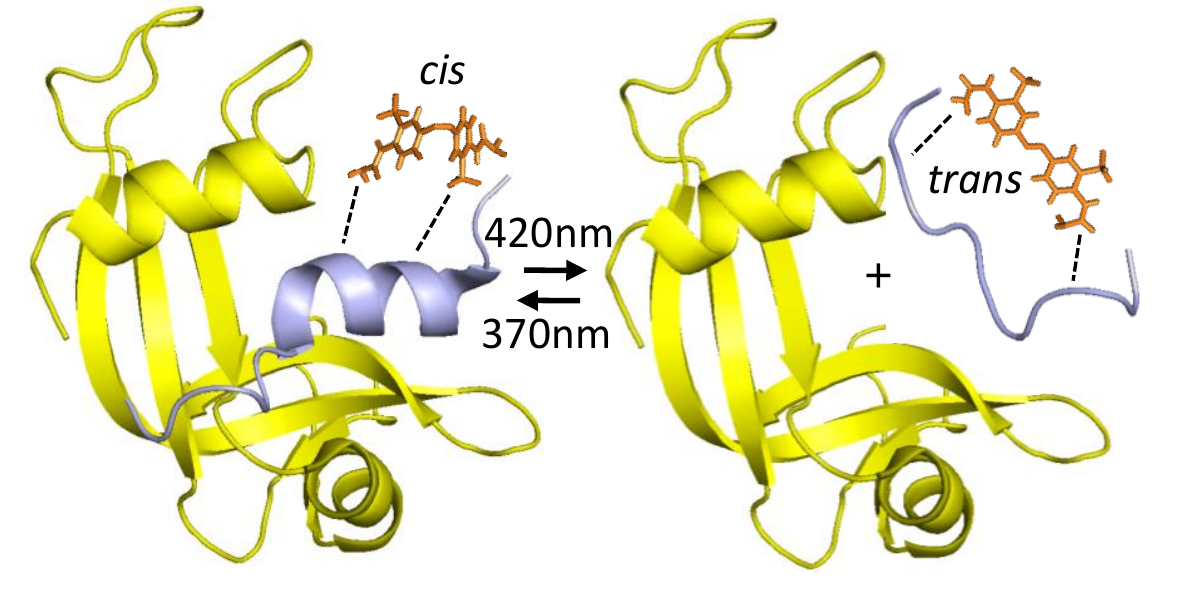}
	\caption{The molecular construct: The S-protein (yellow) binds the photoswitchable S-peptide (grey) in the \textit {cis}-state of the photoswitch (orange). Illumination with 420 nm light promotes the isomerization of the photoswitch to the \textit {trans}-state, initiating the unbinding process.} \label{figStruct}
\end{figure}

\begin{figure*}[t]
	\centering
	\begin{center}
		\includegraphics[width=.8\textwidth]{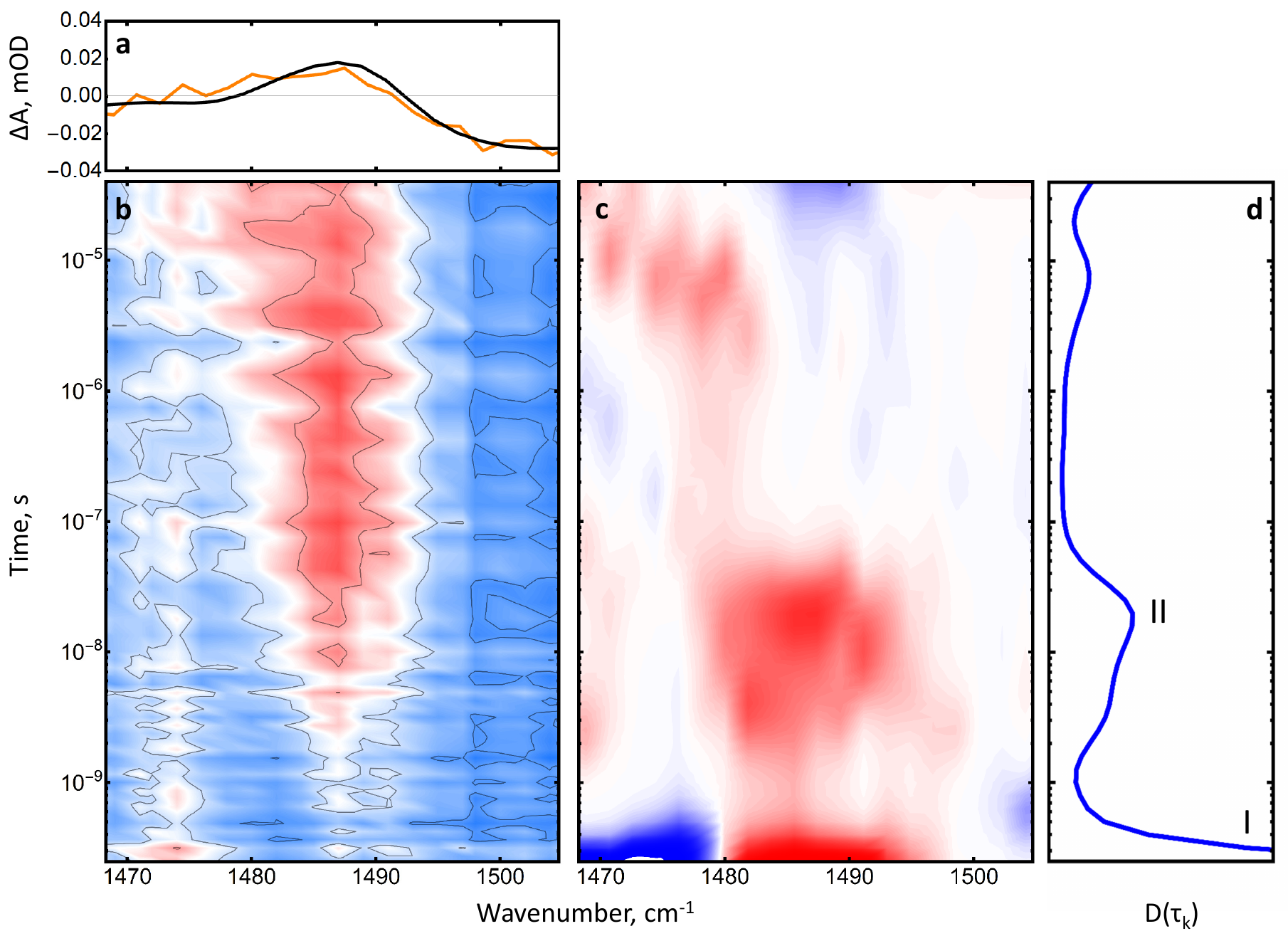}
		\caption{(a) FTIR \textit{cis}-to-\textit{trans} difference spectrum (black) and the last kinetic trace at 42~$\mu$s (orange) of the photoswitch band. (b) Transient IR spectrum of the photoswitch band. (c) Lifetime spectrum Eq.~\ref{EqMultiexp} of the data in panel (b) and (d) its averaged dynamical content Eq.~\ref{Eqdyncontent}.  The two most prominent peaks in the averaged dynamical content, labelled I and II in panel (d), are discussed in the text.}\label{figPhotoswitch}
	\end{center}
\end{figure*}

The free energy landscapes that govern ligand binding are quite shallow, in comparison to the protein folding problem, since the occurring conformational changes are small.\cite{ma1999} This makes the experimental investigation of the protein-protein interactions challenging.\cite{tsai1999} Methods employed to discriminate induced fit from conformational selection are usually based on measuring the binding kinetics and/or on NMR structure determination.\cite{kovermann2017, morando2016, eliezer2007} One example is the NMR-based detailed description of a three-step binding mechanism of an intrinsically disordered domain of the transcription factor (pKID) and the folded protein domain KIX.\cite{sugase2007} The power of NMR methodology lies in providing high structural resolution of the equilibrium ensemble of protein-substates.
A more complete atomistic description of all intermediates and transition states on the pathway from an apo- to a holo-state still remains exclusively the realm of (molecular dynamics (MD) simulations.\cite{robustelli2020, liu2020}

Infrared spectroscopy is sensitive for very subtle conformational and dynamical changes.\cite{Barth2007,fabian2006} In the transient regime, it is the method of choice for the investigation of processes where the time resolution is of paramount importance. To approach the problem of binding from a kinetic perspective, the combination of ultrafast photo-triggering and transient spectroscopy can be employed. One of the most common ways to design photocontrollable proteins/peptides is the crosslinking of an azobenzene moiety at a specific position, whereby its light-induced isomerization leads to certain conformational changes.\cite{mart2016, samanta2011, woolley2005}

RNase S is a non-covalent complex of two fragments of different sizes, the larger S-protein and the smaller S-peptide, which are obtained after cleavage of a single peptide bond of RNase~A.\cite{richards1959} Both fragments bind specifically and form an active complex with in essence the same three-dimensional structure as RNase A, despite the missing peptide bond.\cite{bachmann2011} As a well-studied protein of high stability, RNase S has been used as a model for studying protein-peptide binding mechanisms, among others.\cite{bachmann2011, goldberg1999, schreier1977, luitz2017} The main focus has been the coupled binding and folding of the S-peptide. It is experimentally confirmed that the S-peptide undergoes a disorder-to-order transition, assuming an $\alpha$-helical conformation only in the complex with S-protein, while it is disordered as free peptide.\cite{bachmann2011} More recent MD simulations also provided evidence that the S-protein undergoes substantial conformational changes.\cite{luitz2017}

In this study, we investigate the kinetics of S-peptide unbinding. To that end, we use the previously designed azobenzene-crosslinked S-peptide construct (S-pep(6,13), see Fig.~\ref{figStruct}).\cite{jankovic2019} When the azobenzene photoswitch (orange) is in its \textit {cis}-state, \R{the S-peptide (blue) binds specifically to the S-protein (yellow) by adopting a helical conformation, as has been evidenced by CD spectroscopy.} The S-peptide unbinds completely from the S-protein upon switching to the \textit {trans}-state. Using light-triggered transient fluorescence, we had described unbinding of this peptide as a barrierless process, approaching the speed limit of unbinding.\cite{Jankovic2021}

\begin{figure*}[t!]
	\centering
	\begin{center}
		\includegraphics[width=.8\textwidth]{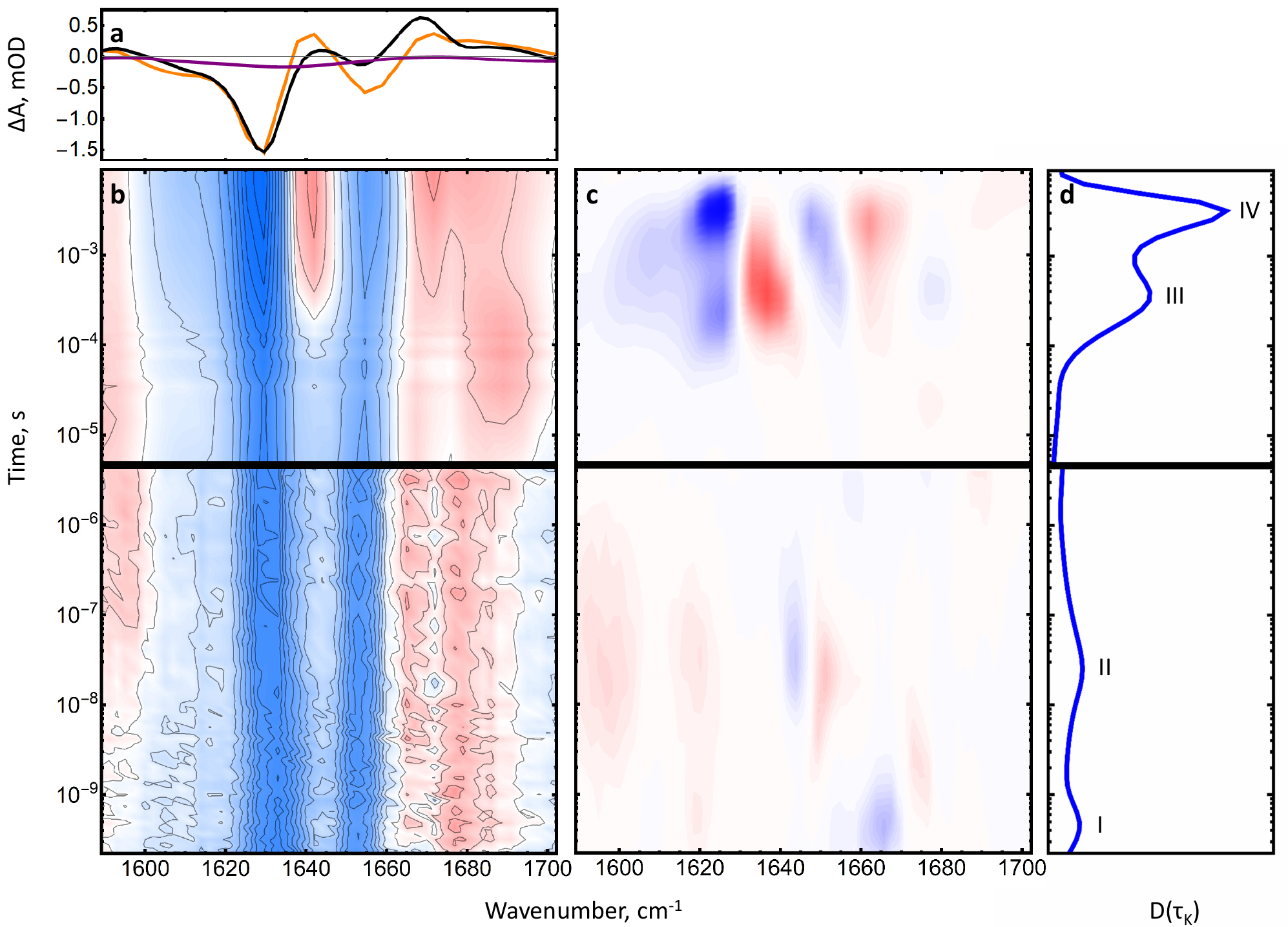}
		\caption{(a) FTIR \textit{cis}-to-\textit{trans} difference spectrum of the complete RNase~S complex (black) and the last kinetic trace at 10~ms (orange) in the amide I region. The purple line shows the FTIR \textit{cis}-to-\textit{trans} difference spectrum of the S-peptide alone for comparison. (b) Transient IR spectrum of the amide I band. (c) Lifetime spectrum Eq.~\ref{EqMultiexp} of the data in panel (b) and (d) its averaged dynamical content Eq.~\ref{Eqdyncontent}. Data acquired with two different experimental setups are stitched together at 5~$\mu$s. Four peaks are identified in the averaged dynamical content, labelled I to IV in panel (d), which are discussed in the text.}\label{figamideI}
	\end{center}
\end{figure*}

Here, we use transient IR spectroscopy to investigate the coupled processes of unbinding and associated conformational changes of the S-protein. %The full process covers more than 7 orders of magnitudes in time, thus providing a detailed mechanistic picture of the protein-peptide interplay  in the RNase S complex.
In these experiments, we start from the \textit{cis}-state of the photoswitch, in which the S-peptide is specifically bound to the S-protein with a binding affinity of $K_d=70~\mu$M.\cite{jankovic2019}  Close to 100\% of the S-peptide is bound at the high concentrations used in our experiments (see Methods). The \textit{cis}-state is prepared with a typical purity of 85\% by pre-illuminating the sample with a 370~nm cw-laser.\cite{bozovic2020a,jankovic2019} Excitation with a short 420~nm (or 447~nm) laser pulse then induces the isomerization of the photoswitch to the \textit {trans}-state. This structural change of the photoswitch eventually leads to unbinding of the S-peptide, as previously confirmed in equilibrium\cite{jankovic2019} and in transient fluorescence experiments.\cite{Jankovic2021}

To reveal the sequence of events that results in S-peptide unbinding, we start with transient IR spectroscopy in the spectral region around 1480~cm$^{-1}$, where we find an isolated vibrational mode that is selective to the configuration of the photoswitch.\cite{Buchli2013} It is in essence the N-D bending mode of the linker between the azobenzene moiety of the photoswitch and the S-peptide backbone. The transient data in Fig.~\ref{figPhotoswitch}b and its lifetime analysis in Figs.~\ref{figPhotoswitch}c and d (see Methods for details) reveals two events, one at a time $<$100~ps (the peak labelled I in Fig.~\ref{figPhotoswitch}d) and the second one around 20~ns (the peak labelled II). The equivalence of the late-time difference spectrum at 42~$\mu$s with an FTIR \textit{cis}-to-\textit{trans} difference spectrum (Fig.~\ref{figPhotoswitch}a), which is at effectively infinite time, evidences that no further process occurs in this spectral region afterwards.

\begin{figure*}[t]
	\centering
	\begin{center}
		\includegraphics[width=0.95\textwidth]{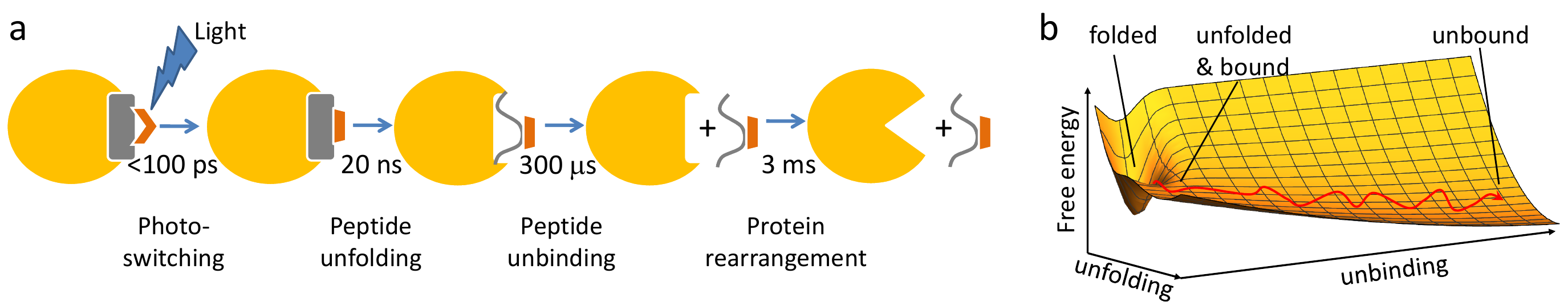}
		\caption{(a) RNase S coupled unbinding and unfolding upon photoswitching the S-peptide conformation reveals 4 major processes and spans more than 7 orders of magnitude in time. (b) \R{Sketch of a free energy surface exemplified for two orthogonal reaction coordinates for S-peptide unfolding and S-peptide unbinding, together with a possible pathway.} }\label{figtimescales}
	\end{center}
\end{figure*}

We attribute the fast process ($<$100~ps, peak I), which we cannot time-resolve with the current experimental setup (see Methods for details), to the actual isomerisation of the azobenzene moiety and the heat it dissipated into its immediate environment.\cite{Hamm1997,Baumann2019} For the subsequent process, we note that the photoswitch is covalently linked to the S-peptide, hence it will be constrained immediately after isomerization, since the RNase~S cannot follow instantaneously. The vibrational mode around 1480~cm$^{-1}$ feels the strain of that linker.\cite{Buchli2013} We attribute the second kinetic component (peak II) to the release of that strain, which is related to the destabilisation of the helical structure of the S-peptide. \R{The observed timescale is in the same range as for the unfolding of a photoswitchable $\alpha$-helix in solution.\cite{ihalainen2007} Furthermore, it has been shown recently by MD simulations that the bound S-peptide is largely $\alpha$-helical when the photoswitch is in the \textit{cis}-state.\cite{jankovic2019} It relaxes into a more stretched unfolded (i.e., higher entropy) conformation, while still residing in the binding pocket, after 10~ns of equilibration in the \textit{trans}-state.}

With that information in mind, we turn in Fig.~\ref{figamideI} to the amide~I spectral region, which covers the C=O stretch vibrations of the backbone of the RNase~S complex and is known to be a sensitive probe of the conformation of peptides and proteins.\cite{barth02} The data were obtained with two experimental setups, covering two different time windows, and are shown together in Fig.~\ref{figamideI} (see Methods for details). The averaged dynamical content  (Fig.~\ref{figamideI}d) reveals 4 dominant processes:
\begin{itemize}

\item Peak I at early times ($<$100~ps), which we attribute to the heat dissipated upon isomerisation of the photoswitch, just like for the photoswitch band in Fig.~\ref{figPhotoswitch}.

\item Peak II around 20~ns, whose timescale is similar to that of the photoswitch band in Fig.~\ref{figPhotoswitch}. We attribute it to unfolding of the $\alpha$-helix, now seen via the amide I band of the S-peptide.

\item Peak III around 300~$\mu$s, which we assign to S-peptide unbinding. This interpretation is based on its timescale, which matches very closely that observed in recent transient fluorescence experiments on the same protein system.\cite{Jankovic2021} The transient fluorescence signal monitors the proximity of the S-peptide to the S-protein, as the azobenzene moiety quenches the fluorescence of the tyrosines of the S-protein, and thus is specifically sensitive to the S-peptide unbinding event.
    %, which has been shown to occur in a stretched exponential manner with a time constant  $\tau$=250~$\mu$s and a stretching factor $\beta=0.5$.

\item Peak IV around 3~ms, which we attribute to the structural response of probably the S-protein after unbinding of the S-peptide. The S-peptide is disordered in the unbound state, and we assume that its structural response is faster. This interpretation is backed up by the observation that the amplitude of the \textit{cis}-to-\textit{trans} FTIR difference spectrum of the S-peptide alone (Fig.~\ref{figamideI}a, purple line) is about 10x smaller compared to that of RNase~S complex (Fig.~\ref{figamideI}a, orange line).

\end{itemize}

The deviation of the late-time transient spectrum measured at 10~ms from the FTIR difference spectrum (Fig.~\ref{figamideI}a, orange vs black) evidences that the process is not completely finished after 10~ms, the time-window of the current experiment. The folding kinetics of RNase~A has been extensively studied, with the slow \textit{cis}-to-\textit{trans} isomerisation of a proline being the rate limiting step.\cite{Udgaonkar1990,Neira1997}

The binding mechanisms of ligands to proteins is often classified as induced fit or conformational selection, while increasing evidence has been provided recently that a combination of both scenarios is possible as well.\cite{boehr2009, morando2016, Bucher2011} The main criterion in this classification is the sequence of events, i.e., the temporal order of binding vs structural modifications of protein and/or the ligand.\cite{boehr2009, Chakraborty2017b} Before discussing our results in the context of these two scenarios, we need to point out two specifics of our experiment: First, in contrast to the binding of a small rigid molecule to a protein, we need to consider that also the S-peptide can undergo significant structural changes. Second, we followed unbinding instead of binding. At a first sight, that seems to contradict the very language of induced fit or conformational selection. However, only in this way we can capture also very fast events, in particular the conformational dynamics of the S-peptide in the binding pocket of the S-protein on a nanosecond time-scale, which would be missed in a binding experiment due to the intrinsically slow and rate-limiting diffusive step. Despite the opposite direction, we can still identify a sequence of events. \R{However, as a word of caution, one needs to keep in mind that the structural processes of binding vs unbinding are not necessarily symmetric.\cite{Okazaki2008}}

Fig.~\ref{figtimescales}a sketches the major events in unbinding of the S-peptide from the S-protein, as documented by transient IR spectroscopy. The overall process covers more than 7 orders of magnitudes in time. Initiated by the isomerisation of the photoswitch, the helical structure of the S-peptide unfolds (20~ns), and precedes the unbinding event (300~$\mu$s) by many orders of magnitudes in time. \R{Here, it is important to note that helix unfolding might be accelerated when driving the process with a photoswitch. For uncoupled $\alpha$-helices in solution, typical timescales in the range of a few 100~ns to 1~$\mu$s have been observed for folding/unfolding.\cite{kubelka04} In addition, for a photoswitchable $\alpha$-helix in solution, it has been observed that the forced unfolding is about 10 times faster (in the range of 100~ns) than folding (in the range of 1~$\mu$s), due to the diffusive search through conformations in the second case.\cite{ihalainen2007} In any case, these timescales are still significantly faster than that of unbinding (300~$\mu$s).}

This result might seem surprising, given that we had concluded in Ref.~\cite{Jankovic2021} that the nominal binding affinity of the S-peptide in the \textit{trans}-state is $K_d=40$~mM, i.e., de facto not existent, \R{and we attributed it to barrierless unbinding. That is, while unbinding is associated with an increase in conformational entropy of the S-peptide, its enthalpic cost is probably minimal due to the special design of the particular RNase~S complex (different from the typical compensation of enthalpy vs entropy known for ligand binding/unbinding).} The peptide nevertheless needs very long time for a conformational search until it eventually unbinds, during which the conformational ensemble in average does not change.
\R{The four processes, photoswitching, S-peptide unfolding, S-peptide unbinding and S-protein relaxation are not independent from each other, despite a clear separation of timescales (in particular for the first three steps). Rather, they form a causal sequence, i.e.,  photoswitching causes S-peptide unfolding, S-peptide unfolding in turn causes S-peptide unbinding, and finally S-peptide unbinding causes S-protein relaxation. In very simple terms, the rate of each step is given by:
\begin{equation}
  k=k_0 e^{-\frac{E_a}{k_B T}}, \label{EqRate}
\end{equation}
where the pre-exponential terms $k_0$ describes a ``search rate'', and $E_a$ an activation energy (which may be zero). However, the reaction coordinate, along which Eq.~\ref{EqRate} is defined, is different for each step, i.e., the various reaction coordinates are orthogonal to each other. Fig.~\ref{figtimescales}b illustrates that idea: after S-peptide unfolding along one reaction coordinate, an orthogonal reaction coordinate becomes relevant for unbinding with a significantly slower search rate $k_0$, indicated by a longer distance of the diffusive search in this direction. This explains why S-peptide unbinding can be orders of magnitudes slower than unfolding, despite the fact that the activation energy for this step is probably very small.
}

These observations speak for an induced fit mechanism, since the S-peptide has a lot of time to sample its conformational space within the binding pocket of the S-protein. Therefore, the coupled unfolding and unbinding proceeds via an induced fit scenario, with unfolding preceding unbinding, similar to other intrinsically disordered systems.\cite{sugase2007,Rogers2014,Rogers2014a,Zou2020} On the other hand, the biggest changes in the S-protein occur only after the unbinding event at 3~ms. This means that the changes in the S-protein are not only induced by the S-peptide, but are happening in the free S-protein as well. Accordingly, from the perspective of the S-protein, conformational selection does describe the process better.

The distinction between conformational selection vs induced fit is a long-standing paradigm in the discussion of ligand binding to a protein. The combination of protein photoswitching and transient IR spectroscopy allowed us to draw direct experimental conclusions about the coupled unbinding and unfolding in RNase S. The complicated interplay of the conformational changes in the S-protein and the intrinsically disordered S-peptide contains elements of both induced fit and conformational selection mechanisms. This result points out the importance of a direct time-dependent structural examination of both partners on all relevant time-scales. We see that a mutually exclusive approach to this problem does not correctly describe the full complexity of the phenomenon. \R{It would be most interesting to look into this processes with the atomistic resolution provided by MD simulations. In this work, we established the relevant timescales, which will help to design a MD setup (that will have to be some form of steered MD simulations).} The same methodology can be applied to other open questions of protein biochemisty, e.g. allosteric communication; whenever access to all transient processes is needed for the full picture to be unraveled.\\

\noindent\textbf{Materials and Methods:} S-protein has been prepared and purified as described previously, with slight modifications described in Ref.\cite{richards1959,Jankovic2021} In brief, commercial ribonuclease A (Sigma-Aldrich) was specifically cleaved by limited proteolysis with subtilisin. The cleavage was performed overnight on ice. S-protein was purified from the mixture of uncleaved ribonuclease A, subtilisin and S-peptide by reverse-phase HPLC on a C5-column and water with TFA and acetonitrile gradient. S-protein eluted at around 40\% acetonitrile. Acetonitrile was removed by size exclusion chromatography. The purity of the final sample was confirmed by mass spectrometry.

S-pep(6,13) was prepared according to the protocol described in Ref.~\cite{zhang03} with slight modifications. Namely, after standard fluorenylmethoxycarbonyl(Fmoc)-based solid-phase peptide-synthesis, the S-peptide (sequence: KETAACKFERQHCDSSTSAA) was dissolved in 50 mM Tris buffer pH 8.5, extensively degassed and reduced with 5 molar equivalents of tris(2-carboxyethyl)phosphine (TCEP). After 30 min of incubation at room temperature, 5 molar equivalents of water-soluble linker molecule (3,3'-bis(sulfonato)-4,4'-bis-(chloroacetamido) azobenzene) was added. The linking reaction was performed overnight at room temperature in the dark. The linked S-peptide was purified using the anionic-exchange chromatography on Q-Trap column (Sigma Aldrich) to remove the free photoswitch and C18 reverse-phase HPLC of the unbound fraction. The linked S-peptide eluted at around 30\% acetonitrile in water. Acetonitrile was removed by evaporation and the final pure peptide was dissolved and dialyzed against double distilled water. Its purity was confirmed by mass spectrometry, while the concentration was determined by amino acid analysis. The final sample was D$_2$O exchanged by lyophilizing and dissolving in 50 mM sodium phosphate buffer pH 7.0 prepared in D$_2$O.

Two experimental setups have been used for the transient IR spectroscopy, covering different time windows. For the time window between $\approx$100~ps and 42~$\mu$s, two electronically synchronized Ti:sapphire laser systems with a repetition rate of \mbox{2.5 kHz} were used.\cite{Bredenbeck2004} The wavelength of the pump laser was tuned to \mbox{840 nm} such that second harmonic generation in a BBO crystal produced pump pulses centered at a wavelength of \mbox{420 nm} to induce \textit{cis}-to-\textit{trans} isomerization of the photoswitch. To avoid the possible sample degradation, the compressing stage after light amplification was bypassed, resulting in stretched pulses of ca. \mbox{60 ps} duration FWHM and a pulse energy of \mbox{3 $\mu$J} focused on \mbox{$\approx$ 140 $\mu$m}. The pump pulses were mechanically chopped at half the repetition rate of the laser setup. The mid-IR probe pulses were obtained in an optical parametric amplifier.\cite{Hamm2000} After the sample, they were passed through a spectrograph and detected in a \mbox{2$\times$64} MCT array detector with a spectral resolution of \mbox{$\approx$2 cm$^{-1}$/pixel}. Multichannel referencing as described in Ref.~\citenum{Feng2017} was used for noise suppression.

For the time window between $5~\mu$s and 10~ms, a Yb-doped fiber laser/amplifier system (short-pulse Tangerine, Amplitude, France) together with an OPA (Twin STARZZ, Fastlite, France) with a subsequent frequency mixing stage in a LGS crystal has been used as a source of femtosecond IR pulses. In this case, a \mbox{2$\times$32} MCT array detector together with a customized measurement electronics\cite{Farrell2020} has been used for detection with a spectral resolution of \mbox{$\approx$4 cm$^{-1}$/pixel}. Data have been taken in a single-VIS-pump-multiple-IR probe fashion,\cite{Greetham2016,Hamm2021} in which case the repetition rate of the laser system defines the time-resolution (10~$\mu$s). VIS pump pulses at 447~nm and with a repetition rate of 10~Hz were generated with a GaN multimode laser diode (PLPT9~450D\_E, Osram Opto Semiconductors), operated by a pulsed laser diode driver (LDP-V~10-10, PicoLAS) to produce pulses of 2~$\mu$s length with a typical pulse energy of 20~$\mu$J.

In either case, S-protein and S-peptide concentrations were 2~mM and 1.5~mM, respectively. With the binding affinity of the peptide in the \textit{cis}-state of $K_d$=70~$\mu$M,\cite{jankovic2019} this resulted in almost 100\% of the peptides bound to the protein. Furthermore, since the protein was in excess, there is basically no unbound peptide, avoiding that the response of free peptide  contaminates that of bound peptides. Free S-proteins in the mixture do not yield any transient response in this experiment, since they are not  modified and do not respond to the pump laser pulse used for photoisomerization of the photoswitch. The sample has been pumped continuously through a cuvette in order to exchange it between subsequent pump laser pulses, using different flow rates for the two
experimental setups in order to adapt to the different repetition rates of the pump lasers.

The timescales contained in a transient data set were determined by fitting each kinetic trace at probe-frequency $\omega_i$ to a multiexponetial function:
\begin{equation}
  S(t,\omega_i) = a_0- \sum_k a(\omega_i,\tau_{k})e^{-t/\tau_{k}} \label{EqMultiexp},
\end{equation}
where a maximum entropy method has been applied for regularisation.\cite{Lorenz-Fonfria2006,Buhrke2020} In this fit, the timescales $\tau_{k}$ were fixed and equally distributed on a logarithmic scale with 10 terms per decade, while the amplitudes $a(\tau_k)$ were the free fitting parameters. The resulting lifetime spectra $a(\omega_i,\tau_k)$ are shown in Figs.~\ref{figPhotoswitch}c and \ref{figamideI}c. The averaged dynamical contents shown in Figs.~\ref{figPhotoswitch}d and \ref{figamideI}d encompass all kinetic processes of the data sets by summing over all probe wavelengths in a way that accounts for the fact that the lifetime spectra $a(\omega_i,\tau_{k})$ can have positive or negative signs:\cite{HammStock2018}
\begin{equation}
  D(\tau_{k})=\sqrt{\sum_i a(\omega_i,\tau_{k})^2}. \label{Eqdyncontent}
\end{equation}
\\

\noindent\textbf{Acknowledgement:} We thank Gerhard Stock and his group for many insightful discussions, Rolf Pfister for the synthesis of the photoswitch molecule, and the Functional Genomics Center Zurich, especially Serge Chesnov, for his work on the mass spectrometry. The work has been supported by the Swiss National Science Foundation (SNF) through the NCCR MUST and Grant 200020B\_188694/1.\\

\noindent\textbf{Data Availability:} The data that support the findings of this study are openly
available in Zenodo (the link will be provided at the proof stage.)\\

\makeatletter
\def\@biblabel#1{(#1)}
\makeatother

\def\bibsection{\section*{}} %removes the horizontal bar before references

\noindent\textbf{References:}
\vspace{-1.5cm}

%\bibliographystyle{achemso}
%\bibliography{references}

\providecommand{\latin}[1]{#1}
\makeatletter
\providecommand{\doi}
  {\begingroup\let\do\@makeother\dospecials
  \catcode`\{=1 \catcode`\}=2 \doi@aux}
\providecommand{\doi@aux}[1]{\endgroup\texttt{#1}}
\makeatother
\providecommand*\mcitethebibliography{\thebibliography}
\csname @ifundefined\endcsname{endmcitethebibliography}
  {\let\endmcitethebibliography\endthebibliography}{}

\end{document}